\documentclass{revtex4-1}
\usepackage[usenames,dvipsnames]{color} \usepackage{ulem} 

\newcommand{\vomega}{\mbox{\boldmath $\omega$}}  
\newcommand{\be}{\begin{equation}}   \newcommand{\ee}{\end{equation}}

\begin{document}        
\title{Helical Fluid and  (Hall)--MHD Turbulence: a Brief Review}
\author{Annick Pouquet$^1$ and Nobumitsu Yokoi$^2$}
 \affiliation{
1- Laboratory for Atmospheric and Space Physics, University of Colorado, Boulder, CO-80303 (USA).
Orcid 0000-0002-2355-2671 \\
2- Institute of Industrial Science, University of Tokyo, Komaba, Meguro, Tokyo 153-8505, Japan. Orcid 0000-0002-5242-7634} 

\begin{abstract} 
Helicity, a measure of the breakage of reflectional symmetry  representing  the topology 
 of turbulent flows,  contributes in a crucial way to their dynamics and to their fundamental statistical properties. We review several of their main features, both new and old, as the discovery of bi-directional cascades or the role of helical vortices in the enhancement  of large-scale magnetic fields in the dynamo problem. 
The dynamical contribution in magnetohydrodynamic (MHD) 
of the cross-correlation between  velocity and induction 
 is discussed  as well. We  consider next how turbulent transport is affected by helical constraints, in particular in the context of magnetic reconnection and fusion plasmas {under one- and two-fluid approximations}. 
 Central issues on how to construct turbulence models for non-reflectionally symmetric  helical
  flows  are reviewed,  including  in the presence of shear, and we finally briefly mention the possible role of helicity in the development of strongly localized  
quasi-singular  structures {at small scale.}   
\\ \\ \\ \\ \\ \\

\large{Paper to appear in {\it Phil. Trans. Roy. Soc.} (2021).} \\

Special issue: {\sl  Scaling the Turbulence Edifice}.
\end{abstract} 

\maketitle         
     

\section{Introduction} \label{S:INTRO}
Lorenz knots, by definition, are generated by the closed periodic orbits of the classical Lorenz attractor. One never stops learning. Of course, one should not. Did you know that all torus knots (the simplest of which being the trefoil knot) are Lorenz knots, as discovered recently by Joan Birman and her collaborators  \cite{birman_13} (see also \cite{abate_12}), thus building a bridge between dynamical systems, chaos and topology? Knots being  one of the  components of helicity,  together with twist and writhe, what is then  the role of helicity (kinetic, magnetic, cross, generalized) in the dynamics of  fluid and (Hall)-MHD turbulence? 
The first discussions of helical structures in turbulent flows were 
{likely}
made in the context of magnetic fields, which are ubiquitous in the universe; they are also often found to be strong when compared to typical velocities, as for example in the galaxy. This would create  an imbalance in the evolution equation-- in the simplest case, the magnetohydrodynamic formulation, were it not for the fact that the Lorentz force exerted by such fields can be rendered weak through alignment of the magnetic induction and the current density. Such force-free fields 
were thus postulated  to exist many years ago, although their origin remained somewhat mysterious. This led Woltjer  \cite{woltjer_58} 
{(see also \cite{elsasser_56})} 
to realize that magnetic helicity $H_M$ 
(definitions and equations are given in \S \ref{S:EQU}) was an invariant of the MHD equations in the absence of forcing and of dissipation, and so was as well the cross helicity (or cross-correlation $H_C$) between the magnetic field and the velocity.

It is only some years later that a similar conservation law was unraveled in the fluid case for the kinetic helicity $H_V$   \cite{moreau_61, moffatt_69, moffatt_92}. 
Helicity corresponds to topological properties of the flows and fields, through knots, links, twists and writhes \cite{moffatt_69, berger_84, kleckner_13}, and their entanglement. Helicity is a pseudo (axial) scalar, since it can change sign upon change of a reference frame from right-handed to left-handed. This corresponds to the (similar) fact that, in a Serret--Frenet frame describing a string in three-dimensional (3D) space, the torsion is pseudo scalar, the line being able to exit the plane formed by the tangent and curvature to the string, upward or downward. 

Helicity appears to play a role in disparate areas of research beyond fluid dynamics and plasmas, from DNA and bio-chemistry to meta-materials \cite{efrati_14}. It is  involved as well in the problem of entanglements of vortex lines in quantum fluids \cite{taylor_16}. Yet another instance is the orientation of the swimming motions of simple biological systems, with possible applications to nanotechnology: chiral behaviour basically allows for more complex 3D motions and thus to follow physical gradients, e.g. of nutrients. Even in an ocean strongly stratified at large scale, at the size of these micro-organisms (of the order of $10 \mu m$), the flow itself indeed is 3D \cite{lancia_19}.

There is a revival of interest in the properties of helical flows, some of  which we discuss here. We  first set-up the stage in \S \ref{S:EQU} before reviewing  briefly  in \S \ref{S:DYN} one of the central role{s} attributed to helicity in its magnetic form, namely the dynamo. \S \ref{S:INV} discusses the recently uncovered properties of helical flows leading to {so-called}
bi-directional cascades and to the link with the existence of sub-invariants, whereas in \S \ref{S:MOD}, we
{deal with} 
 the all-important issue of modelling of such flows, in particular in the presence of an imposed 
{large-scale shear flow.}
 We mention  applications of these concepts to  our close environment, that of the solar wind, and stress the role of the correlation between the velocity and  magnetic fields in altering standard results of transport. Finally, \S \ref{S:CON} presents a brief  conclusion. Several documents can be found useful in going over some of the technical details such as the general form of second- and third-rank tensors in the helical case, issues which will be only briefly mentioned here 
 \cite{monin_79, frisch_95, politano_03}
 (see also \cite{pouquet_19e} for a recent review).

\section{Equations, Definitions and Modelling} \label{S:EQU} 
Let us write the basic equations for the more general case we wish to consider in this paper, that is  
 Hall-MHD, for an  incompressible fluid; ${\bf f}_{u,b}$ forcing terms are included for both the velocity ${\bf u}$ and the magnetic field ${\bf b}$ expressed in terms of an Alfv\'en velocity, with ${\bf b}={\bf B}/\sqrt{\mu_0 \rho_0}\ $):
\begin{eqnarray}
\frac{\partial {\bf u}}{\partial t} =&-&{\bf u} \cdot \nabla {\bf u} - \nabla P + {\bf j} \times {\bf b} + \nu \nabla^2 {\bf u} 
+ {\bf f}_u  ,  \label{eq:hall_mom}\\
\frac{\partial {\bf b}}{\partial t} = &\nabla& \times \left( {\bf u} \times {\bf b} \right) - d_i \nabla \times \left( {\bf j} \times {\bf b} \right) + \eta\nabla^2 {\bf b} + {\bf f}_b  , \label{eq:hall_ind}\\
\ &\nabla& \cdot {\bf u} = 0 \  \ \ , \  \ \ \nabla \cdot {\bf b} = 0 \ .  \label{eq:div}
\end{eqnarray}
 $P$ is the pressure,  ${\bf j} = \nabla \times {\bf b}$ is the current density, ${\bf a}$  the magnetic potential,  with ${\bf b}=\nabla \times {\bf a}$; $\rho_0, \mu_0$  are the  density assumed constant and the vacuum permeability; 
 $\nu$ and $\eta$ are the  kinematic viscosity and magnetic diffusivity. The total energy is the sum of the kinetic and magnetic energy, {\it viz.} $K_T=K_V+K_M=\frac{1}{2}[ |{\bf u}|^2+|{\bf b}|^2]$.
The Hall current is controlled by the   ion inertial length scale $d_i$; the
 MHD equations are obtained for $d_i= 0$. Finally, when we also have ${\bf b}=0, {\bf f}_b=0$, we recover the Navier--Stokes equations for fluid turbulence.
 
The total kinetic, magnetic, cross and hybrid helicities are defined respectively as:
\be H_V= \left< {\bf u} \cdot {\vomega}\right> \ , \ \  H_M = \left< {\bf a} \cdot {\bf b} \right> \ , \ \ H_C= \left< {\bf u} \cdot {\bf b}\right> \ , \ \ 
H_H= H_M +2d_i H_C + d_i^2  H_V \ ,
\ee
 with $\vomega=\nabla \times {\bf u}$ the vorticity ($\langle \cdot \rangle$ denotes ensemble average). 
 We also have $H_H = {\bf V} \cdot \nabla \times {\bf V}$, with 
 ${\bf V} = {\bf a} + d_i {\bf u}$. For $d_i=0$, the hybrid helicity coincides with the magnetic helicity. 
 {Note that $H_V$ is an ideal invariant in the pure fluid case, $H_M,H_C$ in the MHD case and $H_M, H_G$ in Hall MHD.}

{According to the 
 theorem of Emma Noether  \cite{noether_18}, invariance properties are due to symmetries of the underlying equations. For helicity, it is  the relabelling symmetry of Lagrangian particles.
It is at the core of  the invariance of potential vorticity in barotropic flows which, for its nonlinear part, is the cross-correlation (and linkage)  between vorticity and temperature gradients. It is also responsible for the ideal conservation of cross-helicity in MHD \cite{kuroda_90} (see also \cite{padhye_96, fukumoto_08}). 
Finally, in Hall-MHD, it is the dual invariance of the magnetic and hybrid helicities that is linked to this relabelling symmetry, for the ion and for the electron fluids \cite{araki_15} (see \cite{davignon_16} for extended MHD).
}

While the fully isotropic fluid case only necessitates one defining function in terms of 
{a field (velocity)}
 correlation tensor, two are needed in the non mirror-symmetric  case, corresponding in Fourier space to the energy and helicity spectra. This was re-established recently making use of a helical-vector unit system \cite{deussen_20}. It would be of interest to show the equivalence of the second- and third-order tensorial relationships these authors establish with earlier versions of the helical von  {K}\`arm\`an equation \cite{chkhetiani_96}, as well as the helical third-order exact law derived in  \cite{gomez_00, politano_03} (see   \S \ref{S:INV}\ref{SS:EXA})
 { for more details. Thus, we can define the homogeneous  isotropic two-mode two-time correlation function of a fluctuating vector field $\mbox{\boldmath$\chi$}^\prime$ as:
\begin{equation}
	\langle {{\mbox{\boldmath$\chi$}^\prime_i}({\bf{k}};t) \mbox{\boldmath$\chi$}^\prime_j({\bf{k}}';t^\prime)} \rangle
	= D_{ij}({\bf{k}}) Q_\chi(k;t,t^\prime) 
	+ ({i}/{2}) ({k_\ell}/{k^2}) \epsilon_{ij\ell} H_\chi(k;t,t^\prime);
	\label{eq:basic_fld_prop}
\end{equation}
 $D_{ij} = \delta_{ij} - k_i k_j/k^2$ is the projection operator implementing incompressibility, and $Q_\chi, H_\chi$ are the energy and helicity spectral functions of the fluctuations of field $\mbox{\boldmath$\chi$}$. 
 They will be taken as the lowest-order solutions in a modelling decomposition for dynamos, as described below.
Furthermore, note that one can decompose the velocity field on a helical basis $h^\pm({\bf k})$ for each Fourier mode ${\bf k}$ 
(see \cite{waleffe_93, alexakis_17} and references therein); with time omitted, and with $s=\pm$, we have:
  \be {\bf u_k} = u^+{\bf h^+_k} + u^-{\bf h^-_k}   \  \ \  ,  \ \ \ \sqrt{2} {\bf h}^s_k=  {\bf m}_k/|{\bf m}_k| + is {\bf k} \times {\bf m}_k/[| {\bf k} \times {\bf m}|_k] \ \ \ , \ \ \ {\bf m}_k =e_z \times {\bf k} \ , \label{eqhpm} 
   \ee  
where $K^\pm, H^\pm$ will denote their $\pm$ modal energy and helicity.
}

In the pure fluid case (${\bf b}\equiv 0$), Beltrami solutions with $ {\bf u} =  \pm \phi^\prime \vomega $ have been known to exist for long times. They are fully helical, but they 
{eventually become} unstable and turbulence finally develops and decays at the same rate as for random fields,
{if later}. In MHD, Alfv\'en waves, with 
${\bf u}=\pm \phi^{\prime\prime} {\bf b}$ are well-known as well. Note that these Beltrami (force-free) solutions 
have been generalized to Hall-MHD as 
$\nabla \times {\bf j} - \phi {\bf j} + \mu {\bf b}=0$  \cite{mahajan_98}. A related phenomenon is that of  Taylor relaxation \cite{taylor_86} whereby one can minimise the energy while keeping the magnetic helicity constant. Then, through the action of turbulent eddies leading to reconnection of magnetic field lines, as exemplified in fusion configurations, the flow can evolve towards a fully helical state.

{We finish this section by mentioning some of the many modelling techniques that can be used in the context of turbulent flows
at high Reynolds number $Re=UL/\nu$, with ${\bf U}$ the mean field of modulus $U$, and with $L$ a typical large scale.}
Strongly nonlinear and inhomogeneous turbulence lacking reflectional symmetry can be investigated with the aid of the multiple-scale direct-interaction approximation (DIA), which is a combination of the 
{original}
DIA with multiple-scale analysis \cite{yoshizawa_84,yokoi_20}. On the basis of
{ this multiple-scale}
DIA, a propagator [correlation spectrum $Q(k;\tau,\tau')$ and response function $G(k;\tau,\tau')$] renormalisation perturbation expansion, which is suitable for treating strongly nonlinear turbulence at very high Reynolds number \cite{kraichnan_59}, 
{can be}
 performed. In addition, the mean-field inhomogeneity effects, including $u'_j (\partial U_i / \partial x_j)$
	{(with 
	${\bf u}^\prime$ its fluctuation)}
	 are taken into account through the derivative expansion following the introduction of multiple scales 
	 {(with both}
slow and fast variables). The only  assumption in this formulation is that the lowest-order velocity field is homogeneous, isotropic and non-mirror-symmetric, and thus can be written as in equ. (2.6), with 
$\mbox{\boldmath$\chi$}^\prime ={\bf u}^\prime_0$, namely: 
\begin{equation}
	\langle {u'_{0i}({\bf{k}};t) u'_{0j}({\bf{k}}';t')} \rangle
	= D_{ij}({\bf{k}}) Q_0(k;t,t') 
	+ ({i}/{2}) ({k_\ell}/{k^2}) \epsilon_{ij\ell} H_0(k;t,t') \ ;
	\label{eq:basic_fld_prop2}
\end{equation}
here, $Q_0, H_0$ are the energy and helicity spectral functions of the lowest-order field. The first ($Q_0$-related) term represents the mirror symmetric part, and the second ($H_0$-related) term represents the non-mirror-symmetric part. We see from (\ref{eq:basic_fld_prop}) that the breakage of reflectional symmetry enters through the helicity, as expected.
For example, the Reynolds stress is expressed as \cite{yokoi_93}:
\begin{equation}
	\langle {u'_i u'_j} \rangle
	= \langle {u'_{0i} u'_{0j}} \rangle
	+ \langle {u'_{0i} u'_{1j}} \rangle
	+ \langle {u'_{1i} u'_{0j}} \rangle
	= ({2}/{3}) K_V \delta_{ij}
	- \nu_{\rm{T}} {\cal{S}}_{ij} 
	+ [\lambda_{{\rm{H}}i} \Omega_{\ast j}
	+ \lambda_{{\rm{H}}j} \Omega_{\ast i}]_{\rm{D}},
	\label{eq:rey_strss_result}
\end{equation}
where $K_V$ 
{was previously defined.} 
$\mbox{\boldmath${\cal{S}}$} = \{ {\cal{S}}_{ij} \}$ is the rate-of-strain tensor of the mean velocity, 
$\mbox{\boldmath$\Omega$}_\ast = \{ \Omega_{\ast i} \}= \mbox{\boldmath$\Omega$} + 2 \mbox{\boldmath$\omega$}_{\rm{F}}$ the mean absolute vorticity (with 
$\mbox{\boldmath$\Omega$} = \nabla \times {\bf{U}}$ the mean relative vorticity and $\mbox{\boldmath$\omega$}_{\rm{F}}$ the angular velocity). Finally, ${\rm{D}}$ denotes the deviatoric part of a tensor. 
{Two transport coefficients emerge from this formulation: 
a classical turbulent eddy viscosity 
$\nu_{\rm{T}}$ denoting the effect of small scales on large ones, 
and  $\mbox{\boldmath$\lambda$}_{\rm{H}}$ which is related to the helicity-gradient. 
They} 
  are expressed in terms of the spectral functions $Q(k;\tau,\tau_1)$, $H(k;\tau,\tau_1)$ 
  {as well as the response function $G(k;\tau,\tau_1)$, the exact isotropic Green's function obtained from the renormalisation procedure from the bare Green's function $G'_{ij}({\bf{k}};\tau,\tau_1)$ associated with the lowest-order velocity ${\bf{u}}'_0$ equation} as: 
\begin{equation}
	\nu_{\rm{T}} 
	= \frac{7}{15} \int d{\bf{k}} \int_{-\infty}^\tau\!\!\! d\tau_1 
		G(k;\tau,\tau_1) Q(k;\tau,\tau_1),\;\;\;
	\mbox{\boldmath$\lambda$}_{\rm{H}} 
	= \frac{1}{30} \int k^{-2} d{\bf{k}} \int_{-\infty}^\tau\!\!\! d\tau_1 
		G(k;\tau,\tau_1) \nabla H(k;\tau,\tau_1).
	\label{eq:nuT_lambdaH}
\end{equation}
{The equation of $G'_{ij}({\bf{k}};\tau,\tau_1)$ is given as 
\begin{eqnarray}
	\lefteqn{
	\frac{\partial G'_{ij}({\bf{k}};\tau,\tau_1)}{\partial \tau}
	+ \nu k^2 G'_{ij}({\bf{k}};\tau,\tau_1)
	}\nonumber\\
	&&- 2 i M_{i \ell m}({\bf{k}}) \int \int 
	\delta({\bf{k}} - {\bf{p}} - {\bf{q}})
		u'_{0\ell}({\bf{p}}; \tau) G'_{mj}({\bf{q}};\tau,\tau_1) 
		d{\bf{p}} d{\bf{q}}
	= D_{ij}({\bf{k}}) \delta(\tau-\tau_1),
	\label{eq:G_eq}
\end{eqnarray}
where $M_{i\ell m} = [k_m D_{i\ell}({\bf{k}}) + k_\ell D_{im}({\bf{k}})]/2$ and $\delta({\bf{x}})$ is the Dirac's delta function.}

Note that (\ref{eq:rey_strss_result}) (together with (\ref{eq:nuT_lambdaH})) is not heuristically modelled, but is analytically derived from the fundamental equation with the aid of the multiple-scale renormalised perturbation expansion method
{(which of course comes with its own set of hypotheses).}
 Since $G$, $Q$ and $H$ are related to the turbulence time and length scales, which are key quantities for a model, (\ref{eq:rey_strss_result}) and (\ref{eq:nuT_lambdaH}) provide a firm basis for turbulence modelling, 
 {that helps as well guides physical intuition (see also \S~\ref{S:MOD}).
 For instance, the expressions in (\ref{eq:nuT_lambdaH}) provide the information on how the transport coefficients depend on which turbulent statistical quantities are involved, including for the central problem of evaluation of the associated model constants.}
\section{{A brief mention of} helical and non-helical dynamos} \label{S:DYN}
Magnetic induction, like vorticity, is an axial vector, so it may not be surprising after all that kinetic helicity 
{can be}  
involved in the growth of 
{large-scale}
magnetic fields, that is in the kinematic dynamo problem
{for a given velocity field. This phenomenon has been amply reviewed, in terms for example of the so-called alpha-dynamos, in many papers and books
(see  \cite{parker_55, steenbeck_66}, and {\it e.g.} \cite{branden_rev, tobias_21} for references).}
{ Similarly, magnetic helicity is  involved  in the saturation (and possibly the enhancement) mechanism of the large-scale dynamo when the magnetic field is strong enough to react on the velocity \cite{pouquet_76}.}

{More specifically, the}
 generation of magnetic fields by turbulent motions was {described analytically}  more than fifty years ago; 
 it did emphasise  the central role played by kinetic helicity in the large-scale dynamo. Since helicity can be viewed as a specific combination of velocity shear components, one that  remains invariant in the inviscid case, 
{one can remark that helical dynamos in fact  cover as well the case of shear dynamos.}
When using a classical  turbulence model
{(one similar to what was described in the previous section, but Markovianized)},
 one  observes  a substantially different critical magnetic Reynolds number for the onset of dynamo action whether or not the flow is helical  \cite{leorat_81}, a result corroborated by DNS \cite{meneguzzi_81}. It has also been  realized that, even in the absence of helicity but with helicity fluctuations, a dynamo  is still at play \cite{gilbert_88} although it needs a larger separation of scales to be established. 

In fact, the effect on large-scale properties of a turbulent flow in MHD in the helical case was first unraveled with a study of statistical mechanics, taking into account all three invariants ($K_T, H_M, H_C$) \cite{frisch_75}. In that case, there are values of the Lagrange multipliers (corresponding to the invariants) that lead to spectra peaking in the large scales, contrary to the pure hydrodynamic fluid with helicity \cite{kraichnan_73}. It is then not surprising that, in the nonlinear  regime, it is now the invariance of magnetic helicity and its subsequent inverse cascade to large scales which ultimately furnishes a saturation mechanism in combination with Alfv\'en waves \cite{pouquet_76}. Cross-helicity also plays an important role, contributing an extra term to the turbulent electromotive force \cite{yokoi_99, yokoi_13}, a role observed as well in the solar wind \cite{yokoi_06} (see also \cite{zhou_M_90}).

For small-scale dynamos in MHD, again Alfv\'en waves play a central role since they bring into equipartition the (small-scale) kinetic and magnetic energies. Chaos was also seen as determinant 
{in forging such small-scale magnetic fields}
 \cite{galloway_92, ponty_95} (see also \cite{branden_rev}). 
 {Indeed, the}
   enhancement of small-scale magnetic field 
   {has been shown to take}
  place because of  the presence of chaotic streamlines in fully helical (Beltrami) flows for which ${\bf u}= \pm \phi^\prime \vomega$  \cite{dombre_86}. 

The growth of large-scale magnetic energy can be seen as arising from a destabilising mechanism due to small-scale helical properties. On the other hand, for fluids, the turbulent viscosity computed through a Renormalisation Group approach, has been shown to be third-order in an expansion in terms of wavenumbers ($k \rightarrow 0$), and thus sub-dominant to the eddy viscosity \cite{pouquet_78} (see also \cite{zhou_90}).
However, breaking the assumption of isotropy, leads to the so-called anisotropic kinetic alpha (AKA) effect  \cite{frisch_AKA_87}, which is a 
three-dimensional 
large-scale instability of the hydrodynamic flow lacking parity invariance. In the basic procedure deriving the AKA effect, the turbulent Reynolds number is assumed to be small, and the mean velocity is assumed to be uniform over the small-scale length and time. 
In astrophysical and geophysical environments, some range of fluid motions with three-dimensional fluctuations lacking reflectional symmetry may be approximated as independent of the large-scale inhomogeneities. In such a case, the mean velocity evolution can be treated with the AKA large-scale instability
{(see \cite{levina_14} for an application to vertical hot towers and the onset of tropical cyclones).}

\section{The role of invariants on the dynamics}  \label{S:INV}
\subsection{Helical invariants}     
It was realized,  early on in the study of turbulent flows, that invariants, such as the total energy, play a key 
role since interactions between all modes have to sum-up to zero change in the energy in the absence of viscosity and forcing. Thus, the discovery of the invariance of kinetic helicity $H_V$ \cite{moreau_61, moffatt_69} led to the examination of 
its contribution to 
the dynamics of turbulent flows. Written in vector-product form, the Euler equation 
$\partial_t {\bf u} = {\bf u} \times \vomega - \nabla P^\star,$ 
with $P^\star=P+|{\bf u}|^2/2$ the modified pressure, makes the invariance of kinetic helicity particularly straightforward {to derive.}

{Progress in our understanding of helical processes in fluids has been made recently in a variety of ways.
Kinetic}
helicity has been observed in the laboratory \cite{kleckner_13}, making a grid (with a 3D printer) in the form, for example, of a trefoil knot. Such experiments have opened the way to detailed investigations of helical {fluids.} Helicity affects both the processes of vortex twisting  and  of vortex stretching.
{These processes  may differ for flows more complex than Fully Developed Turbulence (FDT) with, for FDT, }
 stronger intermittency and possibly an inverse cascade of energy \cite{yan_20}.
 Direct numerical simulations of reconnection processes {in fluids} were performed for a trefoil knot configuration
\cite{kerr_18} (as well as for anti-parallel vortices), and helicity spectra and reconnection of vortex structures were recently computed using a vortex method \cite{kivotides_21}. 

One would thus expect that, upon the presence of non-zero dissipation and helical forcing at a rate $\tilde \varepsilon_V$, a cascade of helicity towards the small scales would take place, as postulated by Kraichnan on the basis of statistical mechanics \cite{kraichnan_73}. Dimensional analysis immediately gives the 
 Brissaud--Frisch--Lesieur--Mazure (BFLM) 
spectra \cite{brissaud_73} $K_V(k) \sim \tilde \varepsilon_V^{2/3} k^{-7/3} , H_V(k) \sim \tilde \varepsilon_V^{2/3} k^{-4/3}$ (with the helicity transfer rate: $\tilde{\varepsilon}_V = D_t H_V$).
These spectra  are not observed, and are deemed to be impossible in a study relying on a closure of turbulence 
\cite{andre_77}, which also indicates the  inhibition of energy transfer to small scales due to helicity. Rather, helicity behaves as a passive scalar would, with a  cascade  Fourier spectrum of the form, with $\varepsilon_V=D_tK_V$:
\be 
K_V(k)\sim \varepsilon_V^{2/3} k^{-5/3} \ \ ,\ \ H_V(k)\sim \varepsilon_V^{-1/3} \tilde \varepsilon_V k^{-5/3} \ . 
\ee
This allows for a {slow}  recovery of full isotropy at small scales, since $\tilde \sigma_V(k)\equiv H_V(k)/[kK_V(k)]\sim 1/k$, whereas in the BFLM spectrum, we have $\tilde \sigma_V^{(bflm)}(k)=1, \forall k$ in the inertial range
(note that, in the inviscid case, $|\tilde \sigma_V(k)|\sim [k|\kappa/\xi] \le 1$ \cite{kraichnan_73}, where $\kappa, \xi$ are the Lagrangian multipliers appearing in the distribution function). 
{It was also found recently that the angle between Fourier components of velocity and vorticity or induction  decays as $k_\perp^{-1}$ in inertial Alfvenic or fluid turbulence  \cite{milanese_20, milanese_21}.}
Helicity can be defined as well in quantum turbulence \cite{kedia_18}, and this dual cascade of energy and helicity to small scales also occurs in that case, as shown for example using DNS of the Gross--Pitaevskii equations \cite{clark_17}. 
Since energy and helicity are conserved in the inviscid case for any (Fourier) truncation of the system, this includes the minimal set of three modes, or elementary triadic interaction.
This strong conservation property is called detailed balance (see  \cite{moffatt_14} for an analysis of interactions with a small number of triads of wavenumbers). 

From the $\pm$  helical base written in equ. (\ref{eqhpm}), it
 immediately follows that there are three distinguished subsets for which conservation properties hold (together with using the $\pm$ symmetry). For example, the [+/++] subset of three interacting $+$ modes 
 (and also for [-/-\ -]) is globally conservative and has been shown to lead to an inverse cascade of energy \cite{alexakis_18} (see \cite{herbert_E_12} for experimental evidence).
It was shown in \cite{chen_03} using DNS that strong fluxes of the $+$ species, and separately of the $-$ species, 
were observed to almost cancel 
each other as one approached the smallest excited scales of the flow, allowing for a (slow) return to full isotropy at small scale.
Furthermore, when restricting the nonlinear interactions of the Navier--Stokes equations to only one-signed helical modes, it was also shown in \cite{biferale_12} that the helicity cascaded to the small scales while the energy cascaded to the large scales, and in fact, in that case, regularity could be proven, as in two space dimensions
{(2D),} 
for these  bi-directional  constant-flux cascades \cite{biferale_13}.

There are in fact  three *separate* constant fluxes in the helicity cascade  \cite{alexakis_17}.
To our knowledge, this study has not been performed for  MHD, and it would be of interest to do so. It might help unravel the different results concerning cascades of magnetic helicity \cite{mueller_12} as well as in the decay case \cite{ mininni_09}. Indeed, as for kinetic helicity as we mentioned before, it is not the spectrum predicted by dimensional analysis based on the magnetic helicity injection rate that occurs in DNS (although it does so in Markovian closures of MHD \cite{pouquet_76}), but it is rather spectra that are significantly steeper. Magnetic helicity  spectra $H_M(k)\sim k^{-3.3}$ have been observed, and in that case the large scales play a predominant role, with strong {\underline{non-local}} interactions
{between small scales and large scales.}
More investigations of these spectra must be performed
{(see \cite{mininni_08} for the fluid case).}
 Finally note that new helical invariants have also been analyzed in \cite{kelbin_13}.

In the presence of helicity, with broken mirror-symmetry, the $\pm$ degeneracy of the fully isotropic case is lifted and the $K^\pm, H^\pm$ fluxes can be distinguished. An inverse energy cascade, for a subset of  all the possible $\pm$ interactions, is found; it is associated with backscatter and eddy noise although this point needs further investigation.
The other two sub-categories 
{of interactions}
are [+/-+] and [+/-\ -]. Different subsets of these interactions conserve different combinations of $K^\pm, H^\pm$ \cite{alexakis_17}, leading to different sub-dynamics. 
Kraichnan \cite{kraichnan_73} had already noted the possibility that the [+/++] and [-/-\ -] modal subsets could lead to {(partial)}
 inverse energy cascades, but he also remarked that the other 
{remaining}
nonlinear interactions would swamp such effects. 

Moreover, it is  shown in \cite{alexakis_17} that these several fluxes are constant separately in the inertial range.
In fact, the presence of several invariants, beyond global energy and helicity, has also been discussed by a number of authors. For example, the case of helical symmetry, as a generalization of axial symmetry (the axis being now an helix) is discussed in  \cite{deussen_20}. Such flows are, again, quasi two-dimensional (2D); they have an infinite number of invariants, like for the 2D Euler equations, and are proven as well to be integrable \cite{mahalov_90}. 

{\it A priori}, the same {type of argument can be used} for magnetic helicity, and one expects as well sub-invariants that are separately conserved. It is not clear whether this would explain the behavior observed in DNS of MHD for three different flows having the same global invariants ($K_T, H_C, H_M$) and yet settling to different inertial range  scaling \cite{lee_10}. This remains an open question.
Finally, we note that helicity cascades  also occur in rotating turbulence \cite{mininni_09c}. However, in the presence of stratification only, helicity is no more an invariant, and it decays substantially more slowly than the energy \cite{rorai_13},
{contrary to the non-stratified case.}
 When both rotation and stratification are included, as is relevant to the case of the atmosphere and  oceans, helicity can be created 
 {in the quasi-linear regime }
 \cite{hide_02}, as studied in detail in \cite{marino_13h}.

{Finally, it can be shown that, in Hall-MHD, the growth rates of the magnetic and generalized helicity cascades to large scales both vary exponentially with the control parameter, that is the ratio of the forcing scale to the ion inertial scale $d_i$. This result can be recovered using a simple physical argument in which the dimensional scaling law for the helicity spectra, $H_{M,G}(k)  \sim  {\tilde \epsilon}_{M,G}^{2/3} k^{-2}$, plays a central role \cite{pouquet_20}. 
This was  verified by a series of (moderate-resolution) numerical simulations. It would certainly be of interest to pursue this study for substantially higher numerical resolution at small scale, leading to higher Reynolds numbers $Re$, in order to investigate the role small scales play in the development of inverse cascades in that case.  
For example, does the $k^{-2}$ inverse scaling of $H_{M,G}$ persist at high $Re$ 
in Hall-MHD?
}

\subsection{Exact laws in the presence of helicity} \label{SS:EXA}

A conservation law, such as that of kinetic energy, is a global volume-integrated property of non-dissipative physics. These laws are in fact 
{rather} 
constraining in the sense that they are valid, in Fourier space, for each individual (but isolated) triad. They also lead, under a suite of hypotheses (homogeneity, isotropy, stationarity, and non-zero dissipation in the limit of  high Reynolds number), to an exact relationship linking third-order structure functions to the kinetic energy dissipation rate 
$\varepsilon_V$ and the spatial distance $r$ between two points. In the simplest (incompressible) case, this leads to the Kolmogorov 4/5th law \cite{K41b}; with $\delta {\bf u} ({\bf r}) = {\bf u}({\bf x}+{\bf r})- {\bf u}({\bf x})$, and with $u_L$ the longitudinal component of the velocity (along ${\bf r}$), it is written as $ \langle {\delta u_L(r)^3} \rangle = - ({4}/{5}) \varepsilon_V r$.
This law has been generalised to include all components of the velocity,  to the dynamics of the passive scalar and to that of MHD \cite{politano_98g} and Hall-MHD
\cite{galtier_08}. 
In the helical case, some rewriting of second- and third- rank tensors taking into account the fact that vorticity and helicity are pseudo (axial) vectors and scalars, was deemed necessary  
 \cite{oughton_97, politano_03}. 
Vectorial versions of these exact laws stemming from conservation properties can in fact be derived in a simpler manner \cite{banerjee_16}. In the fluid helical case, the exact relationship for helicity reads, with 
$\tilde \varepsilon_V$  the  dissipation rate of $H_V$:
\be \left< \delta u_L({\bf r}) \delta u_i({\bf r}) \delta \omega_i({\bf r}) \right> -  
(1/2)\langle {\delta \omega_L({\bf r}) [\delta u_i({\bf r})]^2} \rangle = - 4 
 \tilde \varepsilon_V r/3   \  \ \ , \ \ \ \tilde \varepsilon_V \equiv D_tH_V \ .
 \ee  
For incompressible helical MHD, working on the magnetic induction equation leads to, with 
$\mbox{\boldmath${\cal{E}}$}^{\rm{turb}}\equiv {\bf u} \times {\bf b}$ the electromotive force (EMF)
(see  \cite{banerjee_16} for Hall-MHD):
\be 
3 \langle {{\large[}\mbox{\boldmath${\cal{E}}$}}^{\rm{turb}}({\bf x}) \times {\bf a}({\bf x}+{\bf r}){\large]}_L  \rangle =  \tilde \varepsilon_{Hm} \ r \  \ \ , \ \ \ 
\tilde \varepsilon_{Hm} \equiv D_t H_M \ .
\ee
Five remarks are in order.
{\bf (i)} The magnetic helicity law is not in terms of structure functions, but of correlations functions, likely because a large-scale magnetic field cannot be eliminated from the dynamics (and is the source of Alfv\'en waves), whereas Galilean invariance allows us to ignore the mean velocity field.
{\bf (ii)}  These exact helical relationships involve cross-correlations, {\it e.g.} between velocity and vorticity
$\mbox{\boldmath$\omega$}$, or the electromotive force and the magnetic potential.
{\bf (iii)} In their vectorial form \cite{banerjee_16}, such laws clearly indicate the relationship between the lack of Beltramisation (non-zero Lamb vector, Lorentz force and Ohm's law) and the amount of transfer through scales.
{\bf (iv)} These helical expressions have barely been analysed on data, with the notable exception of the fluid case for the EDQNM (Eddy Damped Quasi-Normal Markovian) closure  \cite{briard_17}, and for a series of DNS \cite{sahoo_15}. 
It would be of interest to see such analyses carried out for MHD, both for  EDQNM  and for DNS, as well as in the context of the solar wind  for which more refined data is now available, including at small (ionic and even electronic) scales through the recently launched MMS (Magnetospheric Multi-Scale) mission and the Parker Solar Probe.
And finally, 
{\bf (v)}: The direction of the cascade is not determined by these exact laws which can be observed to change sign, for example in the solar wind \cite{sorriso_07}. In fact, helical invariants in extended MHD, which covers both the regimes of Hall current and electron inertia, have a more complex behaviour \cite{kimura-K_14, miloshevich_17}.
$H_H$, which in that paper is called  ion canonical helicity, is shown to undergo either a direct of inverse cascade (in the latter case when the magnetic energy is in excess), a fact that may  be linked to its dominant physical dimensionality as the ion inertial scale varies. This may also be associated with the non-locality of energy transfer, for example in Hall-MHD (see \cite{mininni_07} and references therein).

\subsection{The role of cross helicity} \label{SS:HC}

The cross-helicity $H_C$ is the correlation between the velocity and magnetic fields. It is a conserved volume-averaged correlation directly related to Alfv\'en waves and, in the compressible case, to magnetosonic waves as well \cite{yokoi_18b}.
$H_C$ also plays a role in the estimation of the dissipation and of the magnetic reconnection rate, in particular in shaping the structures within the flow and at its boundaries \cite{yokoi_11, yokoi_11b}. It has been shown numerically to be weak in the reconnection region between two large eddies, allowing it to be effective within the fluid \cite{meneguzzi_96}. The analysis presented in these papers (see also \cite{higashimori_13, yokoi_13b}) gives this result a theoretical basis within a wider modelling framework, and it confirms the finding elaborated  in \cite{grappin_82} (see also \cite{titov_19}) indicating that the development of the cross-helicity spectrum is progressive in time, with positive and negative lobes and finally, at peak of dissipation, with a change of signs pushed to the characteristic dissipation wavenumber. 

For modelling, the large-scale magnetic-field  evolution is subject to a turbulent EMF $\langle {{\bf{u}}' \times {\bf{b}}'} \rangle$. The fluctuating velocity and magnetic field ${\bf u}^\prime, {\bf b}^\prime$
 both depend on the mean velocity shear $\nabla {\bf U}$ as:
$$ {\partial {\bf{u}}' / \partial t + ({\bf{U}} \cdot \nabla) {\bf{u}}' =  \cdots 
- ({\bf{u}}' \cdot \nabla) {\bf{U}} + \cdots \  \ \ , \ \ \ 
\partial {\bf{b}}' / \partial t + ({\bf{U}} \cdot \nabla) {\bf{b}}' = \cdots + ({\bf{b}}' \cdot \nabla) {\bf{U}} + \cdots \ . }$$
 It follows that the equation for $\langle {{\bf{u}}' \times {\bf{b}}'} \rangle$ is subject to the mean velocity shear with the coupling coefficients being the velocity--magnetic-field correlation. This implies that, in the presence of  large-scale velocity inhomogeneities, the turbulent cross helicity may contribute to the large-scale magnetic field. 
 In the multiple-scale DIA framework, the turbulent EMF in the mean magnetic induction equation is expressed as (with $\beta$, $\zeta_R$, $\gamma$ and $\alpha_R$ new transport coefficients \cite{yoshizawa_90,yokoi_13,yokoi_18a}):
\begin{equation}
	\langle {{\bf{u}}' \times {\bf{b}}'} \rangle
	= - (\beta + \zeta_R) \nabla \times {\bf{B}}
	+ \gamma \mbox{\boldmath$\Omega$}_\ast
	+ \alpha_R {\bf{B}}
	- (\nabla \zeta_R) \times {\bf{B}} \ .
	\label{eq:emf_exp}
\end{equation} 
{Thus, the large-scale magnetic-field is subject to a turbulent EMF. The   different 
transport coefficients involve
turbulent statistical quantities based on the fluctuating fields
(turbulent total and residual energies in $\beta,$ and $\zeta_R$, as well as turbulent cross-correlation and residual helicity in $\gamma$ and $\alpha_R$), with specifically: } 
 \be \beta \propto \langle {{\bf{u}}'{}^2 + {\bf{b}}'{}^2} \rangle/2 \ \ , \ \ 
 \zeta_R \propto \langle {{\bf{u}}'{}^2 - {\bf{b}}'{}^2} \rangle/2 \ \  , \  \ 
 \gamma \propto \langle {{\bf{u}}' \cdot {\bf{b}}'} \rangle \ \ , \ \ 
 \alpha_R \propto \langle { - {\bf{u}}' \cdot \mbox{\boldmath$\omega$}' + {\bf{b}}' \cdot {\bf{j}}'} \rangle \ .
 \label{coeff} \ee
{Detailed expressions in the case of strong compressibility are found in \cite{yokoi_18a}.} 
We see 
in (\ref{eq:emf_exp}) that the turbulent cross helicity enters the EMF 
as the coupling coefficient for the mean absolute vorticity $\mbox{\boldmath$\Omega$}_\ast$. The {full} analytical expression for this transport coefficient is 
\begin{equation}
\gamma 
	= \int d{\bf{k}} \int_{-\infty}^{\tau_1} d\tau_1 
		[G_u(k;\tau,\tau_1) + G_b(k;\tau,\tau_1)]  Q_w({\bf{k}};\tau,\tau_1) \ , 
		\label{eq:gamma_exp}
\end{equation}
 where $G_u$ and $G_b$ are the response functions for the velocity and the magnetic field, respectively, and $Q_w$ is the cross-helicity spectral function {(see equ. (\ref{eq:basic_fld_prop}))}. This contribution to the EMF is called the cross-helicity effect in dynamos. It can be an important and very relevant extension of mean-field theory. When implemented in a simple dynamo model constituted by the toroidal and poloidal magnetic-field equations with the EMF (Parker equations), it reproduces an oscillatory behaviour of the mean magnetic field similar to the solar and stellar magnetic-field cycles, even without resorting to a presumed differential rotation profile \cite{yokoi_16b,pipin_18}.

	How much this effect works in a dynamo process depends on how much cross helicity we have. The relative cross helicity (denoted  $\sigma_{\rm{C}}$ in the solar-wind  community) is defined by $H_C$ normalised by the total turbulent  MHD energy:        $\sigma_{\rm{C}} 
	= {{H}_C}/{K_T} 
	= 2 {\langle {{\bf{u}}' \cdot {\bf{b}}'} \rangle} /
		{\langle {{\bf{u}}'{}^2 + {\bf{b}}'{}^2} \rangle}.$ This quantity gives a  dimensionless measure of the importance of  cross-helicity effects. If the turbulent magnetic diffusivity  is mainly balanced by the cross-helicity effect, it is expected that the alignment between the mean electric-current density $\bf{J}$ and the mean absolute vorticity $\mbox{\boldmath$\Omega$}_\ast$ occurs (${\bf{J}} \propto \mbox{\boldmath$\Omega$}_\ast$). This is in marked contrast with the case of balancing between the turbulent magnetic-diffusivity effect and the helicity or $\alpha$ effect, where the mean electric-current density ${\bf{J}}$ configuration parallel to the mean magnetic field ${\bf{B}}$ is realised [force-free state, ${\bf{J}} \propto {\bf{B}}$]. 

Another  important cross-helicity effect 
 is on the momentum transport. The Reynolds and turbulent Maxwell stresses in the mean momentum equation are expressed as \cite{yokoi_13}:
\begin{equation}
	\langle {u'_i u'_j - b'_i b'_j} \rangle
	= ({2}/{3}) K_R \delta_{ij} 
	- \nu_{\rm{T}} {\cal{S}}_{ij}
	+ \nu_{\rm{M}} {\cal{M}}_{ij}
	+ \left[ {
		\lambda_{{\rm{H}}i} \Omega_{\ast j}
		+ \lambda_{{\rm{H}}_j} \Omega_{\ast i}
	} \right]_{\rm{D}},
	\label{eq:rey_turb_max_exp}
\end{equation}
where 
\be K_R = \langle {{\bf{u}}'{}^2 - {\bf{b}}'{}^2} \rangle/2 \ \ , \ \ \nu_{\rm{T}} = (7/5) \beta \ \ , \ \ \nu_{\rm{M}}= (7/5) \gamma \ . \ee
$K_R, \nu_T, \nu_M$ are respectively the turbulent MHD residual energy, and the transport coefficients (eddy diffusivity $\nu_T$,  and 
 cross-helicity-related coefficient $\nu_M$). Furthermore, 
$\mbox{\boldmath$\cal{M}$} = \{ {\cal{M}}_{ij} \}$ is the rate-of-strain tensor of the mean magnetic field, and 
 $\mbox{\boldmath$\lambda$}_{\rm{H}}$ is the helicity-gradient-related coefficient [see equs.~(\ref{eq:nuT_lambdaH}), (\ref{coeff}) and (\ref{eq:gamma_exp})]. The $\nu_{\rm{M}}$-related term implies that the turbulent cross helicity coupled with the mean magnetic strain affects the momentum transport. 
This cross-helicity effect in the momentum transport plays a key role in suppressing the turbulent viscosity and/or inducing a global flow in the solar torsional oscillation \cite{itoh_05}, as well as spontaneous zonal flow generation in fusion plasmas \cite{yoshizawa_99}. It also intervenes in enhancing the magnetic reconnection rate by modulating the outflow configuration \cite{yokoi_13b}. The origin of this effect is reckoned as the vortex-motive force $\langle {{\bf{u}}' \times \mbox{\boldmath$\omega$}'} \rangle$ due to the fluctuating Lorentz force ${\bf{J}} \times {\bf{b}}'$. 


	From the viewpoint of  transport enhancement and suppression, the role of  cross-helicity  is to suppress the enhanced transports 
	(turbulent viscosity $\nu_{\rm{T}}$ and  turbulent magnetic diffusivity $\eta_{\rm{T}} (= \beta + \zeta_R)$)
	 arising from the turbulent energies $K_V$, $K_M$, and $K_T$. In inhomogeneous turbulence where the levels of turbulent energies and helicities vary depending on the production mechanisms due to the mean-field inhomogeneities, the local balance between the transport enhancement and suppression alters in space and time. In the flow domain where the relative cross helicity is large, the effective transport is suppressed. On the other hand, 
	{when $\sigma_C$ is weak},
	the enhanced diffusivity shows a local dominance. This is one of the reasons why the magnetic reconnection rate is expected to be locally drastically enhanced when and where the turbulent cross helicity vanishes,
	{as observed in two-dimensional DNS and models of MHD \cite{pouquet_88, meneguzzi_96}.}

\section{Modelling helical turbulence for fluids and MHD}\label{S:MOD}
	It is known 
	that the standard or simplest   
model with a gradient-transport approximation works poorly in turbulence with cross-flow configurations. For instance, an eddy viscosity  applied to a turbulent swirling pipe flow completely fails to reproduce the 
  axial mean velocity profile near the centreline, as  experimentally observed. This is because the eddy viscosity effect 
  is so strong that an inhomogeneous flow structure is rapidly smeared out. The expression of the velocity correlation (\ref{eq:basic_fld_prop}) implies that the turbulent kinetic helicity and energy 
should be 
statistical quantities for the modelling of non-reflectional symmetric turbulence.
	Of course, one can introduce expansions of the eddy viscosity model that takes into account the role of helicity. For example, in \cite{baerenzung_11}, it was shown that  a helical model performs better than  
when not taking into account helicity 
{correction terms},
also allowing to display the role of rotation in partitioning the large-scale energy cascade.
	
	In (\ref{eq:rey_strss_result}), the eddy-viscosity  $- \nu_{\rm{T}} \mbox{\boldmath${\cal{S}}$}$ is a gradient-diffusion model for  
	momentum transport. On the other hand, the helicity -- or $\mbox{\boldmath$\lambda$}_{\rm{H}}$-related term-- represents a deviation from the gradient-diffusion model. In non-mirror-symmetric turbulence,  
	 helicity being 
	 a structural property of turbulence, should be included in the model expression, in addition to  kinetic energy.
	{One can}  model the transport coefficients $\nu_{\rm{T}}, \mbox{\boldmath$\lambda$}_{\rm{H}}$ (\ref{eq:nuT_lambdaH}), on the basis of the theoretical results, 
{in terms of}
 one-point turbulent statistical quantities ($K_V$, $\varepsilon_V$, and ${H}_V$) as :
 $$\nu_{\rm{T}} = C_\nu \tau K_V = C_\nu K_V^2 /\varepsilon_V \ , \ \mbox{\boldmath$\lambda$}_{\rm{H}} = C_\eta \tau \ell^2 \nabla {H}_V = C_\eta (K_V/\varepsilon_V) (K_V^3/\varepsilon_V^2) \nabla {H}_V \ . $$
	{Such a}
	 turbulence {\it helicity} model with coherent structure effects incorporated through $H_V$ 
	was successfully applied to a turbulent swirling pipe flow \cite{yokoi_93}. 
	 {It thus allowed to unravel}
the role of helicity on coherent structures,
{namely that}
{the presence of turbulent helicity leads to a suppression of turbulent transport, and contributes to a persistent presence of coherent vortical structures against the turbulent mixing due to eddy viscosity.} 
Its systematic extension to  subgrid-scale (SGS) helicity models for large-eddy simulations (LES) 
{has been}  proposed \cite{yokoi_17} 
(also see  \cite{thalabard_15} for a recent attempt to formulate a statistical mechanics framework for large-scale structures of turbulent von K\'{a}rm\'{a}n flows). This helicity effect works 
as well for global structure formation. In a series of 
	DNS in which non-uniform turbulent helicity is injected, 
	a global flow is induced from the initial no-mean flow turbulence in coupling with a system rotation \cite{yokoi_16}. By examining the budget of the Reynolds-stress equation, it was pointed out that the pressure--diffusion correlation as well as the Coriolis-force correlation both play a key role for producing the effect counter-balancing the eddy viscosity \cite{inagaki_17}. Finally, note that the helicity cascade to small scales 
	{has been found to be}
	 slightly less local in scale than its energetic counterpart \cite{yan_20b}. This may lead again to the need of modifying the modelling of helical flows.

In geophysical and astrophysical turbulence with large-scale inhomogeneities, without resorting to any external forcing, helicities are self-generated by these 
 inhomogeneities in combination with rotation or magnetic fields. For  helicity invariants, such as $H_V$ for neutral fluids
 and $H_C$ in MHD, 
 the evolution equations of helicity can be written in a 
 {simplified}
  form for $F = ({H}_V, {H}_C)$, with
 $P_F,\varepsilon_F$ the production and dissipation rates and ${\bf{T}}_F$ the transport flux:
 \begin{equation}
	\partial_t F + ({\bf{U}} \cdot \nabla) F
	= P_F - \varepsilon_F + \nabla \cdot {\bf{T}}_F \ .	\label{eq:turb_evol_eq} \ee
The production rate $P_F$ stems from the mean-field helicity cascade, and is constituted by the inhomogeneity of the mean field coupled with the turbulent flux. For instance, $P_{{H}_V}$ for the turbulent kinetic helicity 
 is expressed as 
	$$P_{{H}_V} 
	= - \langle {u'_i u'_j} \rangle ({\partial \Omega_j}/{\partial x_i})
	+ \Omega_j ({\partial}/{\partial x_i}) 
		\left\langle {u'{}_i u'_j} \right\rangle \ .
	\label{eq:P_HV_def} $$
The dissipation rate $\varepsilon_F$ 
{in equ. (\ref{eq:turb_evol_eq})}
stems from the viscosity and/or diffusivity, representing the helicity decay rate. An elaborate evaluation of $\varepsilon_F$ requires modelling of the  $\varepsilon_F$ equation
{itself,}
 considering the theoretical derivations of the equations of the kinetic helicity dissipation rate $\tilde \varepsilon_V$ and the cross-helicity dissipation rate $\tilde \varepsilon_{HC}$ (see \cite{yokoi_18,yokoi_11}). The transport rate 
 $\nabla \cdot {\bf{T}}_F$
 {in equ. (\ref{eq:turb_evol_eq}),}
  written in the divergent form, represents the flux through the boundary.
Among the several terms in 
$\nabla \cdot {\bf{T}}_F$, the inhomogeneity of turbulence along the angular velocity vector or the mean magnetic field, provides helicity generation through  transport. For ${H}_{V}$, we have $(2 \mbox{\boldmath$\omega$}_{\rm{F}} \cdot \nabla) \langle {{\bf{u}}'{}^2} \rangle/2$, which means that the turbulence inhomogeneity along the rotation direction generates a turbulent kinetic helicity. This is related to an important helicity generation mechanism in rotating stratified turbulence \cite{hide_02, marino_13h}. These production and transport rates, as well as the dissipation rate due to viscosity, constitute the evolution equation of the turbulent helicity.

As we mentioned in \S~\ref{S:INV}\ref{SS:HC}, the cross helicity plays significant roles in the alteration of turbulent flow dynamics. In (\ref{eq:emf_exp}) and (\ref{eq:rey_turb_max_exp}), the $\gamma$- and $\nu_{\rm{M}}$-related terms arise from the turbulent cross helicity. They may counter-balance the terms of the turbulent magnetic diffusivity $\eta_{\rm{T}} (= \beta + \zeta_R)$ and the turbulent viscosity $\nu_{\rm{T}}$. These cross-helicity effects are expected to contribute to the suppression of the turbulent transport in improved-confinement modes in fusion plasmas, and to the localisation of the effective magnetic diffusivity into a small region where the cross helicity vanishes, leading to a fast reconnection in a fully turbulent medium
(see e.g. \cite{meneguzzi_96} in 2D in MHD).
	Another interesting case where the turbulent cross helicity plays an important role is the solar-wind (SW). The  SW and its magnetic field is a dynamically evolving, inhomogeneous and anisotropic turbulent fluid. Since  Alfv\'{e}n waves propagate dominantly outwardly  from the solar surface (the basis of the solar corona), a positive (negative) cross helicity is observed in the inward (outward) heliospheric magnetic sectors. SW turbulence exhibits  strong Alfv\'{e}nicity: the alignment and equipartition between the velocity and magnetic-field fluctuations in the inner heliosphere and the Alfv\'{e}nicity both decay as the heliocentric distance increases.
	
	 Turbulence models for the large-scale evolution of the solar wind have been developed \cite{zhou_M_90,tu_93}. In these models, the evolution of turbulence 
	 using 
	  the  kinetic and magnetic energies $K_V, K_M$
	(or equivalently, the  total and residual energies $K_T =K_V + K_M, K_R=K_V-K_M$),
{as well as}
 the turbulent cross helicity ${H}_C$, 
has been investigated. Characteristics of the large-scale behaviour of the Alfv\'{e}nicity represented by the normalised cross helicity 
$\sigma_{\rm{C}}$ 
were obtained 
 with the aid of  spacecraft observations \cite{roberts_90}. They 
 were successfully reproduced by a turbulence model based on the multiple-scale DIA analysis of  MHD turbulence \cite{yokoi_06,yokoi_07,yokoi_08}.

Indeed, high-resolution spacecraft observations of  solar-wind turbulence provide important data for constructing models of  energy cascades in Hall-MHD  and beyond, as for instance statistical theories of extended MHD where both the Hall drift and the electron inertia are included to treat length and time scales comparable to the ion cyclotron frequency and/or the electron skin depth \cite{miloshevich_17,miloshevich_20} (see also \cite{zhu_14}). These formulations are expected to be useful
for modelling  inertial confinement by high-intensity lasers in the laboratory, as well as turbulence--plasma waves interactions. 
For example, a recent comparison of  high-resolution MMS spacecraft data with the analysis of 2D hybrid-kinetic numerical simulations confirmed that  kinetic contributions to inter-scale energy transfer can remain important at  scales  larger than  the ion-inertial length \cite{bandyopadhyay_20}, possibly because of the inverse (magnetic  and generalized) helicity cascades, {and for intermittency constrainsts as well.}

Finally, it will be important in the future to incorporate the mechanisms behind bi-directional cascades in statistical theories because of their role in the mixing and dissipation properties of turbulent flows
(see \cite{marino_15p} {in the oceanographic context for a model taking into account the slowing-down of turbulence transfer in the presence of waves combined with the dominance of rotation at large scale).}
 
\section{Conclusions}  \label{S:CON}
Helicity is an integrand part of {fluid dynamics}, including in the possibility of the development  of singular structures,
{a central tenet of turbulence research leading, among other, to intermittency and to anomalous (order unity) dissipation for the high Reynolds number flows encountered in geo- and astrophysics.}
 Indeed, it was shown for fluids that, if a singularity (in the sense of a finite-time blow-up of vorticity) occurs, then the curvature of vortex lines blows up as well \cite{constantin_08}; but what of torsion, the third element in the Serret--Frenet frame, which can be associated with helicity? This may well be an open problem, and similarly for MHD. In fact, the role of helicity in such singular structures was emphasised again recently  \cite{migdal_20} 
 {in the framework of the Clebsch variable formalism.}
  Similarly, one could ask what role, if any, helicity might have for turbulence in non-integer dimension, studied in \cite{frisch_12} using a fractal Fourier decimation, possibly {\it e.g.} with the help of fractional derivatives \cite{shlesinger_93}.

Moreover shear, an integral part of the definition of 
helicity, has been shown recently, for rotating and/or stratified fluids as well as for (Hall)-MHD turbulence, to have a central role  in the formation of strong {small-scale} localized  structures, {an interpretation of which can be performed for example}
in the framework of critical phenomena
\cite{farrell_12, shih_16, barkley_16}. {The observation of such shear instabilities} in complex turbulent flows  affects our evaluation of mixing and dissipation in the presence of  rotation and stratification \cite{fritts_09, feraco_18}, {with local values comparable to that of FDT. It also
leads for example to high (intermittent) vertical drafts in the mesosphere \cite{chau_21}  (see \cite{klymak_18} for oceanic observations).}
{This is occurring as well in the presence of  magnetic fields, in which case cross-helicity plays a central role \cite{sorriso_19}, as already elaborated in this review. This in fact can be related to an interpretation in terms of  classical avalanches  \cite{pouquet_19p, smyth_19}, as developed earlier  to understand the distribution spectrum of the energy of solar flares \cite{bak_87, lu_91}. However,  the role of the breaking of reflectional symmetry and of helical structures in general in the establishment of so-called self-organized criticality for such cases has not been entirely elucidated to this date.
}

Finally, detailed observations of magnetic helicity in solar flares have already allowed for the prediction of  strong solar eruptions  \cite{pariat_17}. In addition to various physics-based models including the helicity-dependent or non-parity-invariance effect on turbulent transport \cite{frisch_AKA_87,baerenzung_11,yokoi_93}, using highly resolved data sets from detailed laboratory experiments, high-performance computing, high-resolution in-situ satellite observations, and analysing  and combining them with a variety of {modelling and of}
machine learning techniques \cite{raissi_19, taghizadeh_20},  all this is expected to lead to enhanced physical insights in turbulence and nonlinear phenomena  in general. 

\vskip0.12truein
{\it We appreciated useful remarks made by the referees.
Annick Pouquet is thankful to LASP, and in particular to Bob Ergun. }
Nobumitsu Yokoi was partly supported by Japan Society for the Promotion of Science (JSPS) Grants-in-Aid for Scientific Research 18H01212. No direct data analysis is performed herein, and we have no competing interests. 


 \end{document}